\documentclass[12pt]{article}
\usepackage{epsfig}

\def\beq{\begin{equation}}
\def\eeq#1{\label{#1}\end{equation}}
\def\eeqn{\end{equation}}

\def\beqa{\begin{eqnarray}}
\def\eeqa#1{\label{#1}\end{eqnarray}}
\def\eeqan{\end{eqnarray}}

\let\bar=\overbar

\def\Dslash{\not{\hbox{\kern-4pt $D$}}}
\def\dslash{\not{\hbox{\kern-2pt $\del$}}}

\def\msb{{\bar{\ssstyle M \kern -1pt S}}}

\def\Title#1{\begin{center} {\Large {\bf #1} } \end{center}}

\begin{document}

\Title{Results on $D^+$ $\to \ell^+ \nu$ and $D^+_{s} $ $\to
\ell^+ \nu$ decays at Charm factories}

\bigskip\bigskip


\begin{raggedright}

{\it L.L. Jiang\index{Reggiano, D.}\\
Institute of High Energy Physics\\
No.918 P.O.X,Beijing,China}\\

\bigskip\bigskip
\end{raggedright}

\noindent
Proceedings of CKM 2012, the 7th International Workshop on the CKM Unitarity Triangle, University of Cincinnati, USA, 28 September - 2 October 2012

\section{Introduction}

   In the Standard Model (SM) of particle physics, $D^+_{(s)}$\footnote{throughout this
manuscript, charge conjugation is implied} mesons can decay into
$\ell^+\nu$ (where $\ell^+$ is $e$, $\mu$ or $\tau$) through
a virtual $W^+$ boson. The virtual $W^+$ boson is produced in the
annihilation of the $c$ and $\overline d$($\overline s$) quarks. The
decay rate of this process is determined by the wave-function
overlap of the two quarks at the origin, and is parameterized by the
$D^+_{(S)}$ decay constant, $f_{D^+_{(S)}}$.
Figure 1 shows the decay diagram for the
Cabibbo-suppressed purely leptonic decay of the $D^+_{(S)}$ meson.
In this decay process, all strong interaction effects between the two quarks in initial state
are absorbed into $f_{D^+_{(S)}}$.
\begin{figure}[htpt]
\centerline{
\includegraphics[width=10.0cm,height=4.0cm]{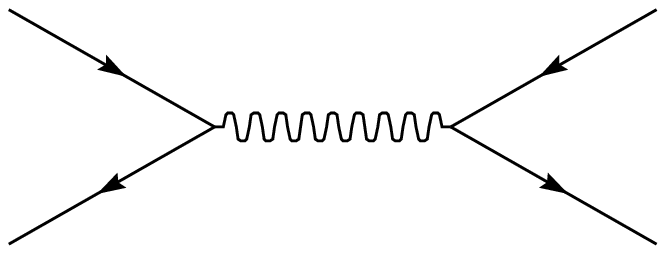}
\put(-282,50){\large\bf $D^+_{(S)}$}
\put(-20,100){\large\bf $\ell^+$}
\put(-20,17){\large\bf $\nu$}
\put(-265,100){\large\bf $c$}
\put(-255,10){\large\bf $\bar d (\bar s)$}
\put(-155,70){\large\bf $W^+$}
           }
\caption{The decay diagram for $D^+_{(S)} \rightarrow \ell^+\nu$.}
\label{feynman_digram}
\end{figure}
The decay width of $D^+_{(S)} \rightarrow \ell^+\nu$
is given by the formula~\cite{quarks_and_leptons}
\begin{equation}
\Gamma(D^+_{(S)} \rightarrow \ell^+\nu)=
     \frac{G^2_F f^2_{D^+_{(S)}}} {8\pi}
      \mid V_{cd(s)} \mid^2
      m^2_{\ell} m_{D^+_{(S)}}
    \left (1- \frac{m^2_{\ell}}{m^2_{D^+_{(S)}}}\right )^2,
\label{eq01}
\end{equation}
where $G_F$ is the Fermi coupling constant, $V_{cd(s)}$ is the
$c\rightarrow \bar d(\bar s)$ Cabibbo-Kobayashi-Maskawa (CKM) matrix
element~\cite{pdg2010}, $m_{\ell}$ is the mass of the lepton, and
$m_{D^+_{(S)}}$ is the mass of the $D^+_{(S)}$ meson.
The ratios of decay rates for $D^+ \rightarrow \tau^+\nu_{\tau}$, $D^+ \rightarrow \mu^+\nu_{\mu}$ and
$D^+ \rightarrow e^+\nu_{e}$ are expected by the well-known
lepton-masses, which are
\begin{eqnarray}
R &=&  \Gamma(D^+ \rightarrow \tau^+\nu_{\tau}) : \Gamma(D^+ \rightarrow \mu^+\nu_{\mu}) : \Gamma(D^+ \rightarrow e^+\nu_{e})
    \nonumber \\
  &=& 2.67 : 1 : 2.35 \times 10^{-5}.
\label{eq01}
\end{eqnarray}
while the ratios of decay rates for $D^+_s \rightarrow \tau^+\nu_{\tau}$, $D^+_s \rightarrow \mu^+\nu_{\mu}$ and
$D^+_s \rightarrow e^+\nu_{e}$ are expected to be 9.8:1:2.4$\times 10^{-5}$. Any significant deviation of experimental
results from theory expectation may indicate that there are New Physics (NP).

The pseudoscalar decay constants $f_{D^+_{(s)}}$ are very important constants
in heavy flavor physics.
Within the context of the SM, measurement of the purely leptonic
decay of the $D^+_{(s)}$ meson provides a means of determining $f_{D^+_{(s)}}$.
As the decay constant is related to the annihilation probability of the heavy and the light
quarks inside the meson, they play an important role both in characterizing the properties
of confinement and as absolute normalization of numerous heavy-flavor transitions, including
semileptonic decays and non-leptonic decays of the mesons
as well as mixing of neutral and anti-neutral meson pairs.
For example, $f_B$ relates the measurements of the $B\bar{B}$ mixing ratio to CKM matrix
elements.
At present it is impossible to precisely determine $f_{B^+}$ experimentally from the purely
leptonic $B^+$ decay and is never possible to measure $f_{B^0_s}$ since $B^0_s$ does not have charge current
leptonic decay, so theoretical calculations of the $f_{B^+}$ and$f_{B^0_s}$ have to be used.
The ratios of $f_{D^+}/f_{B^+}$ and $f_{D_s^+}/f_{B^0_s}$ from LQCD calculations are determined with higher precision
than the calculations of each of the decay constants.
The LQCD calculations of $f_B$ can be used with some level of confidence given by the stringent tests from leptonic D decays on $f_{D^+}$ and $f_{D_s^+}$.

Conversely, taking the theoretical calculations of the decay constants as input, one can
also determine the CKM matrix elements $|V_{\rm cd}|$ and $|V_{\rm cs}|$ by analyzing the $D^+$
and $D^+_{s}$ purely leptonic decays.  Thanks to the unquenched LQCD calculations of
$f_{D^+}$, which have reached a high precision of ~2\%, one can get more precisely experimental result of
$|V_{\rm cd}|$ than the one historically measured based on the $D \to \pi \ell^+ \nu$ semileptonic
decays.

\section{Methods}

The $D$ and $\bar D$ meson are produced in pairs near
open-charm meson pair production energy
thresholds in $e^+e^-$ experiments, which provides the cleanest experimental
environment for studies of the leptonic decays of $D^+_{(s)}$ meson.
Taking advantage of the $D \bar D$ production, if a $D$ meson
decay is fully reconstructed (this is called singly tagged D meson),
a $\bar D$ must exist in the system recoiling against the tagged
$D$ meson. Based on the accumulated singly tagged $D$ meson sample, one can
measure the absolute branching fraction of the leptonic decays of $D^+_{(s)}$.
The neutrino in the final states of leptonic decays of $D^+_{(s)}$
can be reconstructed with the missing momentum and the missing energy of the events.
If there is a neutrino in the recoil side of the singly tagged $D$ meson, the distribution
of the missing mass squared, which is the missing energy square minus the missing momentum square,
should characterize with a peak at zero.

\section{Leptonic decays of $D^+$ meson}

Historically MARK-III, BES-I, BES-II and CLEO-c experiments made measurements of $f_{D^+_{(s)}}$.
Today's running BES-III experiment focus on measurement of $f_{D^+}$.

\subsection{Results at previous experiments}

\subsubsection{Search for $D^+ \rightarrow l^+\nu$ decay at Mark-III experiment}

In 1988, MARK III collaboration first searched for the decay of $D^+ \rightarrow \ell^+\nu$
and got no signal event.
They set an upper limit on the decay constant at $90\%$ C.L. to be
$f_{D^+} < 290$ MeV.~\cite{mark-iii_fD}.

\subsubsection{First observed one candidate for $D^+ \rightarrow \ell^+\nu$ at the BES-I experiments}

In 1998, the BES collaboration observed one candidate event for $D^+ \rightarrow \ell^+\nu$
in the recoil side of the 10082 singly tagged $D$ mesons, by analyzing
22.3 pb$^{-1}$ of data taken at 4.03 GeV. From this experiment, they
measured the branching fraction for $D^+ \rightarrow \mu^+\nu$ to be
$(0.08^{+0.16+0.05}_{-0.05-0.02})\%$, corresponding to a value of
decay constant of $f_{D^+}=(300^{+180+80}_{-150-40})$ MeV~\cite{bes-i_fD}.

\subsubsection{First absolute measurements of $B(D^+ \rightarrow l^+\nu)$ and  $f_{D^+}$ at the BES-II experiments}

At the ``04 Electroweak Interactions $\&$ Unified Theories" international meeting, the BES-II
collaboration reported their analysis results of the leptonic decay of $D^+$ for the first time.
With 33 pb$^{-1}$ of data taken in $e^+e^-$ annihilation with
their upgraded BES-II detector at the BEPC collider, they accumulated
$5321\pm 149\pm 160$ $D^-$ single tags and observed 3 candidates for $D^+ \rightarrow \mu^+\nu$ decays
in the recoil side of the single tags. With the 3 candidates, which contain 0.3 background events estimated from MC simulation
and the accumulated $5321\pm 149\pm 160$ $D^-$ single tags, they measured the
branching fraction for $D^+ \rightarrow \mu^+\nu$ decays to be
$B(D^+ \rightarrow \mu^+\nu)=(0.122^{+0.111}_{-0.053}\pm 0.010)\%$, corresponding to
a value of the decay constant $f_{D^+}=(371^{+129}_{-119}\pm 25)$ MeV~\cite{bes-ii_fD}.
Their analysis results were finally published in 2005.
These are absolute measurements of the decay branching fraction and decay constant, which do
not depend on the yield of $D^+$ meson production and do not depend on some branching fractions
for $D^+$ meson decay into other modes.

\subsubsection{Precise measurements of $B(D^+ \rightarrow l^+\nu)$ and  $f_{D^+}$ at CLEO-c experiments}

CLEO-c published 3 paper to report their studies of the
leptonic decay of $D^+$ mesons in 2004~\cite{cleo-c_fD_2004}, 2005~\cite{cleo-c_fD_2005} and 2008~\cite{cleo-c_fD_2008}, respectively. Because the
previous results are superseded by the most recent one, we here only discuss the results
published in 2008. In that paper, the CLEO collaboration analyzed 818 pb$^{-1}$ of data taken at 3.773 GeV.
They accumulated $460055\pm 787$ $D^-$ tags with 6 hadronic decay modes
of the $D^-$ meson and observed $149.7 \pm 12.0$ signal events for $D^+ \rightarrow \mu^+\nu$.
In the abstract and the summary in that paper, they presented the measured the decay branching fraction of
$B(D^+ \rightarrow \mu^+\nu)=(3.82 \pm 0.32 \pm 0.09)\times 10^{-4}$ and the corresponding
decay constant of  $f_{D^+}=(205.8 \pm 8.5 \pm 2.5)$ MeV~\cite{cleo-c_fD_2008}.

\subsection{New results at BES-III experiment}

The BES-III collaboration has collected 2.9 fb$^{-1}$ of data at 3.773 GeV with
BES-III detector~\cite{bes3} at the BEPC-II~\cite{bepc2} during the time period from 2010 to 2011.
In this section, we report measurements of
the branching fraction for $D^+\rightarrow \mu^+\nu_{\mu}$ decay and
the pseudoscalar decay constant $f_{D^+}$
obtained by analyzing this data sample.

\begin{figure}
\centerline{
\includegraphics[width=12.0cm,height=8.0cm]{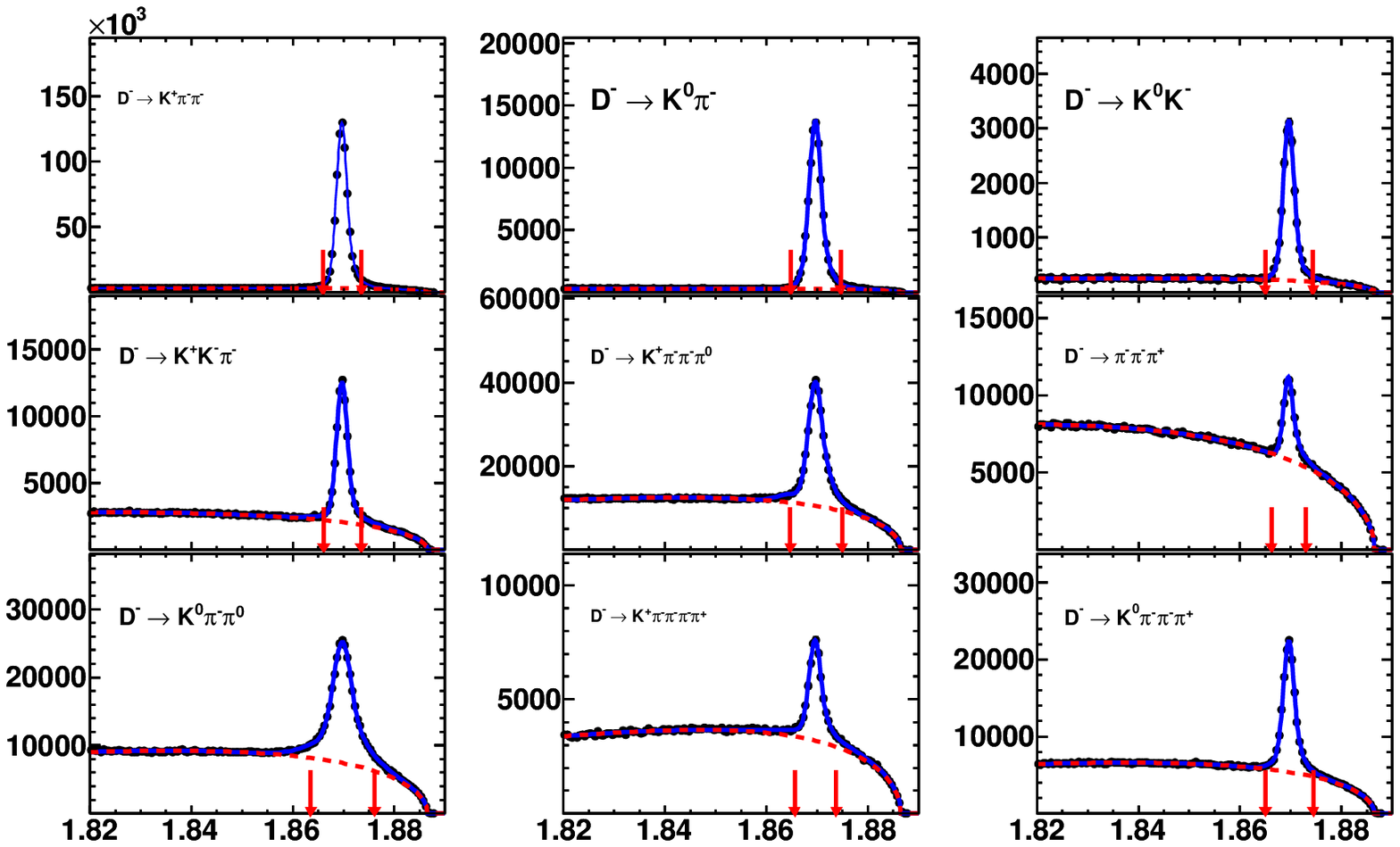}
\put(-212.0,5.0){{M$_{\rm B}$ ~~[GeV/$c^2$]}}
\put(-350.0,80){\rotatebox{90}{Number of Events}}
\put(-300.0,160.0){(a)} \put(-195.0,160.0){(b)} \put(-95.0,
160.0){(c)} \put(-300.0,100.0){(d)} \put(-195.0,100.0){(e)}
\put(-95.0, 100.0){(f)} \put(-300.0,50.0){(g)}
\put(-195.0,50.0){(h)} \put(-95.0, 50.0){(i)}
        }
\caption{Distributions of the beam energy constraint masses of the
$mKn\pi$ combinations for the 9 single tag modes from the data;
where (a), (b), (c), (d), (e), (f), (g), (h), (i) are for the modes
of $D^- \rightarrow K^+\pi^-\pi^-$, $D^- \rightarrow K^0_s\pi^-$,
$D^- \rightarrow K^0_s K^-$, $D^- \rightarrow K^+K^-\pi^-$, $D^-
\rightarrow K^+\pi^-\pi^-\pi^0$, $D^- \rightarrow \pi^+\pi^-\pi^-$,
$D^- \rightarrow K^0_s\pi^-\pi^0$, $D^- \rightarrow
K^+\pi^-\pi^-\pi^-\pi^+$, and $D^- \rightarrow
K^0_s\pi^-\pi^-\pi^+$, respectively. } \label{fig3}
\end{figure}

Nine non-leptonic $D^-$ decay modes
$K^+\pi^-\pi^-$,
$K^0_s\pi^-$,
$K^0_s K^-$,
$K^+K^-\pi^-$,
$K^+\pi^-\pi^-\pi^0$,
$\pi^+\pi^-\pi^-$,
$K^0_s\pi^-\pi^0$,
$K^+\pi^-\pi^-\pi^-\pi^+$,
and
$K^0_s\pi^-\pi^-\pi^+$ are used in accumulating the singly tagged $D^-$ mesons.
The singly tagged $D^-$ mesons are fully reconstructed by requiring
that the energy of the reconstructed candidate $D^-$ must be consistent with the
beam energy,
and then examine the beam energy constraint mass
of the tagged $mKn\pi$ system,
\begin{equation}
M_{\rm B} = \sqrt {E_{\rm beam}^2-|\vec {p}_{mKn\pi}|^2},
\end{equation}
where $E_{\rm beam}$ is the beam energy,
and
$|\vec p_{mKn\pi}|$ is the magnitude of the momentum
of the daughter particle $mKn\pi$ system.
The $M_{\rm B}$ distributions and the maximum likelihood fit results for the $M_{\rm B}$ distributions
of the nine $D^-$ tag modes are shown in Fig.~\ref{fig3}.
The fits give a total of $1586056 \pm 2327$ $D^-$ tags. After subtracting 20103 double counting
$D^-$ tags, BES-III reconstructed $1565953 \pm 2327$ $D^-$ tags which are used
for further analysis of measuring
the branching fraction for $D^+ \rightarrow \mu^+\nu_{\mu}$ decays.

Candidate events for the decay $D^+ \rightarrow \mu^+\nu_{\mu}$
are selected from the surviving charged tracks in the system recoiling against the
singly tagged $D^-$ mesons.
For the $D^+ \rightarrow \mu^+ \nu_{\mu}$ final states,
there should be only one candidate charged track, which is identified as muon.
At BES-III, the $\mu^+$ can be well identified with the Muon detector (MUC).
Taking the difference of the passage lengths in MUC for difference types of the charged
tracks, the candidate muon is required to have large passage length.
Except for this, no extra good photon with energy greater than 300 MeV is allowed to be present.

\begin{figure}
\centerline{
\includegraphics[width=12.0cm,height=11.0cm]{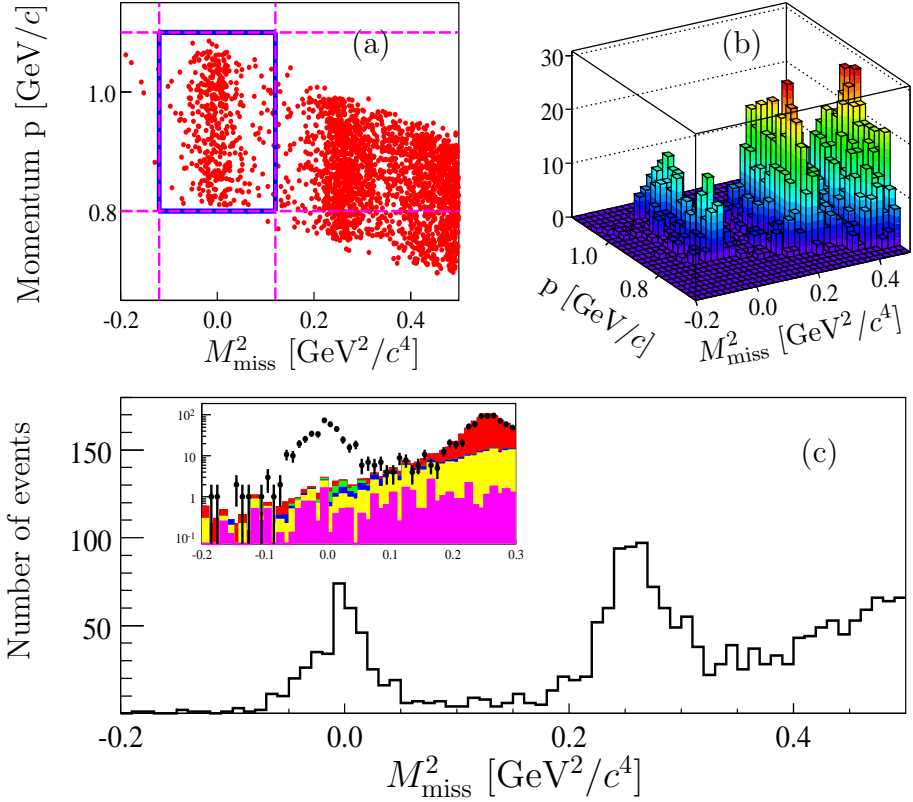}
\put(-276.0,158.0){$M^2_{\rm miss}$~[GeV$^2$/$c^4$]}
\put(-347.0,180){\rotatebox{90}{Momentum p [GeV/$c$]}}
\put(-90.0,158.0){\rotatebox{10}{\small $M^2_{\rm miss}$~[GeV$^2$/$c^4$]}}
\put(-150.0,186){\rotatebox{-32}{\small p [GeV/$c$] }}
\put(-207.0,-3.0){\large{$M^2_{\rm miss}$~[GeV$^2$/$c^4$]}}
\put(-350.0,45){\rotatebox{90}{Number of events}}
\put(-220.0,275.0){(a)}
\put(-80.0,275.0){(b)}
\put(-50.0, 120.0){(c)}
           }
\caption{Distributions of $M^2_{\rm miss}$, where (a) and (b) are scatter plots
of the identified muon momentum $p$ VS $M^2_{\rm miss}$,
and (c) is the distribution of $M^2_{\rm miss}$.
The insert shows the signal region for $D^+ \rightarrow \mu^+\nu_{\mu}$
on a log scale, where dots with error bars are for the data,
histograms are for the simulated backgrounds from
$D^+ \rightarrow K^0_L \pi^+$ (red),
$D^+ \rightarrow \pi^0\pi^+$ (green),
$D^+ \rightarrow \tau^+ \nu_{\tau}$ (blue) and
other decays of $D$ mesons (yellow) as well as
from $e^+e^- \rightarrow$non-$D\bar D$ decays (pink).
}
\label{pmu_vs_umiss_bes3}
\end{figure}
Figures~\ref{pmu_vs_umiss_bes3}(a) and (b) show the scatter-plots of
the momentum of the identified muon satisfying the requirement for
selecting $D^+\rightarrow \mu^+\nu_{\mu}$ decay versus $M^2_{miss}$,
where the blue box in Fig.~\ref{pmu_vs_umiss_bes3}(a) shows the
signal region for $D^+\rightarrow \mu^+\nu_{\mu}$ decays. Within the
signal region, there are $425$ candidate events for $D^+ \rightarrow
\mu^+\nu_{\mu}$ decay. The two concentrated clusters out side of the
signal region are from $D^+$ non-leptonic decays and some other
background events. The events whose peak is around 0.25
GeV$^2$/$c^4$ in $M^2_{miss}$ are mainly from $D^+ \rightarrow
K_L^0\pi^+$ decays, where $K_L^0$ is missing. Projecting the events
for which the identified muon momentum being in the range from 0.8
to 1.1 GeV/$c$ onto the horizontal scale yields the $M^2_{miss}$
distribution as shown in Fig.~\ref{pmu_vs_umiss_bes3}(c). From this
plot, we can see that the difficultly suppressed backgrounds from
$D^+ \rightarrow K_L^0\pi^+$ decays in CLEO-c
measurement~\cite{cleo-c_fD_2008} are effectively suppressed due to
that they use the MUC measurements to identify the muon.
Detailed Monte Carlo studies show that there are
$47.7 \pm 2.3 \pm 1.3$
background events in $425$ candidates
for $D^+ \rightarrow \mu^+\nu_{\mu}$ decays,
where the first error is the MC statistical error and second is the systematic, which
arises from uncertainties in the branching fractions 
or production cross sections for the source modes.
After subtracting the number of background events,
$377.3\pm 20.6 \pm 2.6$ signal events
for $D^+ \rightarrow \mu^+\nu_{\mu}$ decay are retained,
where the first error is statistical and the second systematic arising from the uncertainty
of the background estimation.

    The overall efficiency for observing the decay ${D^+\rightarrow \mu^+\nu_{\mu}}$
is obtained by analyzing full MC simulated events
of $D^+\rightarrow \mu^+\nu_{\mu}$ versus $D^-$ tags
and combining with $\mu^+$ reconstruction efficiency of the MUC.
The $\mu^+$ reconstruction efficiency of the MUC is measured with
muon samples selected from the same data taken at 3.773 GeV. The overall efficiency is
$0.6382~\pm 0.0015$.

   With $1565953$ singly tagged $D^-$ mesons, $377.3\pm 20.6 \pm 2.6$
$D^+ \rightarrow \mu^+\nu_{\mu}$ decay events and with a reconstruction efficiency of
$0.6382~\pm 0.0015$,
the BES-III collaboration obtain the branching fraction
$$B(D^+ \to \mu^+\nu_{\mu})=(3.74 \pm 0.21 \pm 0.06)\times 10^{-4}~~({\rm BESIII~Preliminary}), $$
where the first error is statistical and the second systematic.
This measured branching fraction is consistent within error with
world average of $B(D^+ \to \mu^+\nu_{\mu})=(3.82 \pm 0.33)\times 10^{-4}$~\cite{pdg2010},
but with better precision.

    The decay constant $f_{D^+}$ can  be obtained
by inserting the measured branching fraction, the mass of the muon,
the mass of the $D^+$ meson, the CKM matrix element
$|V_{\rm cd}|= 0.2252\pm0.0007$ from the CKMFitter~\cite{pdg2010},
$G_F$
and the lifetime of the $D^+$ meson~\cite{pdg2010}
Eq.(\ref{eq01}), which yields
$$f_{D^+} = (203.91 \pm 5.72 \pm 1.97)~~\rm MeV~~({\rm BESIII~Preliminary}),$$
\noindent
where the first error is statistical and the second is systematic, which arises
mainly from the uncertainties in
the measured branching fraction ($1.7\%$),
the CKM matrix element $|V_{\rm cd}|$ ($0.3 \%$),
and the lifetime of the $D^+$ meson ($0.7\%$)~\cite{pdg2010}.
The total systematic error is $1.0\%$.

\section{Leptonic decays of $D_s^+$ meson}

The BES-I and CLEO-c experiments have published results on $f_{D_s^+}$
and  BES-III  will publish their experiment results of the purely leptonic decay of $D^+_s$ meson soon.

\subsection{BES-I experiment near $D_s^+D_s^-$ threshold}

The first absolute measurements of decay branching factions of $D_s^+\rightarrow \mu^+\nu$,
$D_s^+\rightarrow \tau^+\nu$ and the decay constant were made by
BES-I experiment in 1995. They found 3 events of both
the $D_s^+\rightarrow \tau^+\nu$ and $D_s^+\rightarrow \mu^+\nu$ decays in the recoil side of
$94.3\pm 12.5$ singly tagged $D_s^-$ mesons by analyzing 22.3 pb$^{-1}$ of data taken at 4.03 GeV.
They measured the decay branching fractions of
$B(D_s^+\rightarrow \tau^+\nu)=(15^{+13+3}_{-6-2})\%$ and
$B(D_s^+\rightarrow \mu^+\nu)=(1.5^{+1.3+0.3}_{-0.6-0.2})\%$, respectively, and the
decay constant of
$f_{D_s^+}=(430^{+150}_{-130} \pm 40)$ MeV~\cite{bes-i_fDs}.

\subsection{CLEO-c experiment near $D_s^+D_s^{*-}$ threshold}

CLEO-c collected 600 pb$^{-1}$ of data at 4.17 GeV in  $e^+e^-$ annihilation. With this
data sample, they published 3 papers to present the studies of the purely
leptonic decays of $D^+_s$ by reconstructing the $D^+_s$ with various decay modes \cite{cleo-c_fDs_2009a, cleo-c_fDs_2009b, cleo-c_fDs_2009c}.
By averaging the results obtained by analyzing different leptonic decay modes of $D^+_s$ meson,
CLEO-c gave $f_{D_s^+}=(259.0 \pm 6.2 \pm 3.0)$ MeV~\cite{cleo-c_fDs_2009c}.

\section{Summary}

Since the first attempt to search for the $D^+$ leptonic decay by the MARK-III experiment in 1988,
many experiments have been making great efforts to search for and study the $D^+$ and $D_s^+$ leptonic decays.
At present, the most precise measurements of the decay constant of $f_{D^+}$ is made by BES-III, which was first reported
at Charm2012~\cite{rongg_charm2012}. They measured the branching fraction $B(D^+ \rightarrow \mu^+\nu)=(3.74 \pm 0.21 \pm 0.06)\times 10^{-4}$,
$f_{D^+}=(203.9\pm 5.7 \pm 2.0)$ MeV and $|V_{\rm cd}|=(0.222 \pm 0.006 \pm 0.005)$.
In addition, more precise results on $D^+ \to \tau^+\nu_{\tau}$, $e^+\nu_{e}$ will be determined from the full BES-III data
taken at 3.773 GeV in the near future, and results on $D^+_s \to \tau^+\nu_{\tau}$, $\mu^+\nu_{\mu}$ will be reported from the BES-III recent selected
data at 4.010 GeV soon.

\section*{Acknowledgement}
I would like to thank the colleagues of BES-III collaboration,¡¡the staff of BEPCII and¡¡the computing
center of IHEP for their hard efforts and excellent works.
This work is partly supported by National Key Basic Research Program (973 by MOST) (2009CB825200) and¡¡
National Natural Science Foundation of
China (NSFC) under Contract No. 10935007.

\end{document}